\documentclass[twocolumn,superscriptaddress,nofootinbib,amsmath,amssymb,preprintnumbers,longbibliography]{revtex4-2}

\usepackage{bm}
\usepackage{graphicx}
\usepackage{hyperref}
\usepackage{mathtools}
\usepackage{color}
\usepackage{physics}

\allowdisplaybreaks[1]

\hypersetup{
colorlinks = true,
linkcolor = blue,
citecolor = blue,
anchorcolor = blue,
urlcolor = blue
}

\newcommand\+{\dagger}
\newcommand\<{\langle}
\renewcommand\>{\rangle}

\renewcommand\Re{\mathrm{Re}}
\renewcommand\Im{\mathrm{Im}}

\newcommand\J{{\bm{J}}}

\newcommand\p{{\bm{p}}}
\newcommand\q{{\bm{q}}}
\renewcommand\k{{\bm{k}}}
\newcommand\0{\mathbf{0}}
\renewcommand\r{{\bm{r}}}

\newcommand\up{\uparrow}
\newcommand\down{\downarrow}

\begin{document}
\preprint{RIKEN-iTHEMS-Report-23}
\title{Spin conductivity spectrum and spin superfluidity in a binary Bose mixture}

\author{Yuta Sekino}
\affiliation{RIKEN Cluster for Pioneering Research (CPR), Astrophysical Big Bang Laboratory (ABBL), Wako, Saitama, 351-0198 Japan}
\affiliation{Interdisciplinary Theoretical and Mathematical Sciences Program (iTHEMS), RIKEN, Wako, Saitama 351-0198, Japan}
\author{Hiroyuki Tajima}
\affiliation{Department of Physics, Graduate School of Science, The University of Tokyo, Tokyo 113-0033, Japan}
\author{Shun Uchino}
\affiliation{Advanced Science Research Center, Japan Atomic Energy Agency, Tokai, Ibaraki 319-1195, Japan}

\date{\today}
\begin{abstract}
We investigate the spectrum of spin conductivity for a miscible two-component Bose-Einstein condensate (BEC) that exhibits spin superfluidity.
By using the Bogoliubov theory, the regular part being the spin conductivity at finite ac frequency and the spin Drude weight characterizing the delta-function peak at zero frequency are analytically computed.
We demonstrate that the spectrum exhibits a power-law behavior at low frequency, reflecting gapless density and spin modes specific to the binary BEC.
At the phase transition points into immiscible and quantum-droplet states, the change in quasiparticle dispersion relations modifies the power law.
In addition, the spin Drude weight becomes finite, indicating zero spin resistivity due to spin superfluidity.
Our results also suggest that the Andreev-Bashkin drag density is accessible by measuring the spin conductivity spectrum.
\end{abstract}

\maketitle

\section{\label{sec:introduction}Introduction}
Transport of spin has been extensively studied in various subfields of physics.
In solids, the spin Hall effect~\cite{sinova2015spin} plays a central role in the context of spintronics~\cite{zutic2004spintronics}.
Generation of spin currents by the spin-vorticity coupling has been observed in liquid mercury~\cite{takahashi2016spin}.
A similar phenomenon has also been reported for heavy ion collisions in nuclear physics~\cite{starcollaboration2017global}.
In atomic physics, the advent of ultracold atoms has opened up a precious opportunity to investigate physics of spin transport because one can widely tune various parameters of the systems, control spatial configuration of spin, and observe dynamics of spin-resolved densities~\cite{gross2017quantum,enss2019universal,Krinner2017two-terminal}.

Recently, the present authors have suggested that ac spin conductivity called optical spin conductivity is accessible with existing methods of cold-atom experiments~\cite{sekino2022optical},
while its measurement in condensed-matter materials is demanding.
This quantity would be a valuable probe for various quantum states of matter because it is the spin counterpart of optical conductivity, which plays a crucial role to experimentally investigate exotic electron systems such as high-$T_\mathrm{c}$ superconductors~\cite{homes1993optical} and graphene~\cite{mak2008measurement}.
In addition, understanding of ac spin transport with cold atoms may provide guidelines for application of ac spin currents in spintronics.

Compared to the optical mass conductivity~\cite{tokuno2011spectroscopy,wu2015probing,anderson2019conductivity}, the optical spin conductivity has advantages to examine ultracold atomic gases in which disorder effects are 
usually negligible~\cite{sekino2022optical,enss2012quantum}.
(Note that, in the case of charge neutral atoms, transport of mass or particle number corresponds to charge transport in electron systems.)
In the cases of harmonically trapped or homogeneous systems,
the dependence of the optical mass conductivity on ac frequency is completely determined at the algebraic level by the generalized Kohn's theorem~\cite{wu2015probing,kohn1961cyclotron,brey1989optical,li1991electrodynamic}.
As a result, the conductivity spectrum is independent of details of many-body states including interatomic interactions and is useless as a probe.
Therefore, the presence of an optical lattice or impurity potentials is essential to probe cold-atom systems by the mass conductivity~\cite{anderson2019conductivity}.
On the other hand, a spin current is generally affected by interactions between particles of different spin components even in the absence of lattices or disorder~\cite{enss2012quantum,sommer2011universal}.
Therefore, the optical spin conductivity is sensitive to details of quantum many-body states and works as a probe for both continuum and lattice systems.

So far, several theoretical studies show that the optical spin conductivity provides information on a wide range of physical quantities: a spin drag coefficient, Tan's contact, superfluid gap, and quasiparticle excitations in a Fermi gas~\cite{enss2012quantum,sekino2022optical}, spin excitations and quantum corrections in a spinor Bose-Einstein condensate (BEC)~\cite{sekino2022optical}, a Tomonaga-Luttinger liquid parameter of spin in one-dimensional systems~\cite{sekino2022optical}, and topological phase transition~\cite{tajima2022optical}.

Here, we consider a binary mixture of BECs, where two components are regarded as spin-up and spin-down, and theoretically investigate the spectrum of its optical spin conductivity.
Unlike in the previous cases, this Bose mixture is known to exhibit unique spin dynamics due to spontaneous symmetry breaking associated with spin degree of freedom~\cite{pethick_smith_2008,pitaevskii2003,kasamatsu2005vortices}.
We focus on a $\mathbb{Z}_2$ symmetric mixture, which is invariant under the exchange of spin-up and spin-down atoms.
Such a mixture is realized with $^{23}$Na atoms~\cite{Bienaime2016spin-dipole} and considered as a spin superfluid~\cite{flayac2013superfluidity}, whose evidences have been reported by recent experiments~\cite{fava2018observation,kim2020obvervation}.

In addition, we wish to shed light on the Andreev-Bashkin effect~\cite{andreev1976three} in the Bose mixture, where a superflow of one component drives the mass current of the other without dissipation.
The Andreev-Bashkin effect is a beyond-mean-field phenomenon and therefore has stirred up many theoretical studies~\cite{fil2005nondissipative,linder2009calculatioin,nespolo2017andreev,sellin2018superfluid,utesov2018effective,pasiri2018spin,karle2019coupled,romito2021linear,carlini2021spin,Ota2020thermodynamics,Contessi2021Collisionless}.
It is also known that this nondissipative phenomenon affects various spin dynamics such as a spin sound wave and spin dipole oscillation~\cite{fil2005nondissipative,romito2021linear}.
The Andreev-Bashkin effect has also been discussed in the contexts of liquid $^3$He \cite{volovik2022vortices}, superconductors~\cite{dean1993supercurrent}, and the mixture of superconducting protons and superfluid neutrons in neutron stars~\cite{Alpar1984rapid}.

In this paper, we clarify how the spectrum of the spin conductivity reflects excitation properties, in particular, two gapless modes arising from spontaneous symmetry breaking.
We then investigate the impact of quantum phase transitions to immiscible~\cite{pitaevskii2003} and quantum-droplet phases~\cite{petrov2015quantum,cabrera2018quantum} on the spectrum.
Furthermore, we compute the spin Drude weight, which characterizes the delta function peak at zero frequency in the optical spin conductivity.
The spin Drude weight is found to take a finite value, indicating that the mixture has a zero spin resistivity intrinsic to a spin superfluid.
We also connect the optical spin conductivity with the Andreev-Bashkin effect.

This paper is organized as follows:
In Sec.~\ref{sec:model}, we provide the model of the binary mixture of BECs and the analysis of the optical spin conductivity within the Bogoliubov theory.
Sections~\ref{sec:regular} is devoted to investigations of the regular part of the spin conductivity at finite frequency.
In Sec.~\ref{sec:Drude}, we discuss the spin Drude weight and its connection with the Andreev-Bashkin effect.
We conclude in Sec.~\ref{sec:conclusion}.
In this paper, we set $k_\mathrm{B}=\hbar=1$.

\section{\label{sec:model}Model}
We consider a homogeneous miscible two-component BEC in which interatomic interactions are solely characterized with $s$-wave scattering lengths~\cite{pitaevskii2003}.
The two components are referred to as spin-up ($\tau=\ \up$) and spin-down ($\tau=\ \down$) states.
We focus on a $\mathbb{Z}_2$ symmetric mixture, where both components have the same particle number $N_\up=N_\down=N/2$, mass $m_\up=m_\down=m$, and intracomponent scattering length $a_{\up\up}=a_{\down\down}=a>0$.
In the miscible phase, the intercomponent scattering length $a_{\up\down}$ satisfies $|a_{\up\down}|<a$, while $a_{\up\down}=a$ and $a_{\up\down}=-a$ correspond to the quantum phase transition points to an immiscible mixture and a quantum droplet, respectively.

The Hamiltonian of the binary mixture is given by~\cite{pitaevskii2003}
\begin{align}\label{eq:H}
H&=\sum_{\tau=\up,\down}\sum_{\k}\epsilon_{k}a_{\k,\tau}^\+a_{\k,\tau}\cr
&\quad+\frac{g}{2V}\sum_{\tau=\up,\down}\sum_{\k_1\k_2\p}a_{\k_1+\p,\tau}^\+a^\+_{\k_2-\p,\tau}a_{\k_2,\tau}a_{\k_1,\tau}\cr
&\quad+\frac{g_{\up\down}}{V}\sum_{\k_1\k_2\p}a_{\k_1+\p,\up}^\+a^\+_{\k_2-\p,\down}a_{\k_2,\down}a_{\k_1,\up},
\end{align}
where $\epsilon_{k}=\k^2/(2m)$ is the kinetic energy, $a_{\k,\tau}$ is the annihilation operator of bosons with momentum $\k$ and spin $\tau$, $g=\frac{4\pi a}{m}$ and $g_{\up\down}=\frac{4\pi a_{\up\down}}{m}$ are intra- and intercomponent coupling constants, respectively, and $V$ is the volume.
To investigate this system, we employ the Bogoliubov theory applicable to a weakly interacting mixture at low temperature $T$.
Within the Bogoliubov theory, $a_{\0,\tau}$ is replaced with $\sqrt{N/2}$, and operators with nonzero momenta in Eq.~\eqref{eq:H} are retained up to the quadratic terms.
The resulting $H$ can be diagonalized by using combination of unitary and conventional Bogoliubov transformations given by
\begin{align}\label{eq:b}
\begin{pmatrix}
a_{\k,\up}\\a_{\k,\down}
\end{pmatrix}
=
\frac{1}{\sqrt{2}}
\begin{pmatrix}
1&1\\1&-1
\end{pmatrix}
\begin{pmatrix}
u_{k,d}b_{\k,d}-v_{k,d}b_{-\k,d}^\+\\
u_{k,s}b_{\k,s}-v_{k,s}b_{-\k,s}^\+
\end{pmatrix},
\end{align}
where $b_{\k,d}$ and $b_{\k,s}$ are annihilation operators of quasiparticles associated with density and spin fluctuations, respectively.
For convenience, a new label $\alpha=d,s$ describing density and spin modes and new couplings $g_d=\frac{g+g_{\up\down}}{2}$ and $g_s=\frac{g-g_{\up\down}}{2}$ are defined.
The coefficients in Eq.~\eqref{eq:b} are given by $u_{k,\alpha}=\sqrt{\frac12\left(\frac{\epsilon_k+g_\alpha n}{E_{k,\alpha}}+1\right)}$ and $v_{k,\alpha}=\sqrt{\frac12\left(\frac{\epsilon_k+g_\alpha n}{E_{k,\alpha}}-1\right)}$, where $n=N/V$ is the number density and
\begin{align}\label{eq:u_v_E}
E_{k,\alpha}=E_{\alpha}(\epsilon_k)=\sqrt{\epsilon_k(\epsilon_k+2g_\alpha n)}
\end{align}
is the excitation energy of a quasiparticle in the $\alpha$ mode with momentum $\k$.
The diagonalized $H$ has the form of
\begin{align}\label{eq:H_2}
H=\sum_{\alpha=d,s}\sum_{\k\neq\0}E_{k,\alpha}b_{\k,\alpha}^\+b_{\k,\alpha},
\end{align}
where the ground state energy irrelevant to spin transport was ignored.
In the presence of the intercomponent interaction ($g_{\up\down}\neq0$), the two modes are not degenerate ($E_{k,d}\neq E_{k,s}$).

For the sake of simplicity, we throughout focus on the situation where the depletion of the condensate is negligibly small.
As shown in Appendix~\ref{appendix:depletion}, this condition is satisfied when the gas parameters $\sqrt{na_\alpha^3}\ll1$ are sufficiently small and temperature $T$ is very low compared to $\mathrm{max}\{g_d n,\,g_s n\}$ and $T_\mathrm{BEC}^0$, where $T_\mathrm{BEC}^0=2\pi(n/[2\zeta(3/2)])^{2/3}/m$ is the BEC transition temperature for a binary mixture of noninteracting bosons and $\zeta(z)$ is the zeta function.
Therefore, the condensate density $n_0$ can be approximated as $n_0\approx n$.

\subsection{\label{sec:optical_spin_conductivity}Optical spin conductivity}
In this section, we evaluate the optical spin conductivity of a binary mixture of BECs.
Following our previous paper~\cite{sekino2022optical}, we consider the mixture in the presence of a weak spin-driving force $\bm{f}_{s}(t)$ to induce an ac spin current.
In ultracold atomic gases, such an external force can be generated by a magnetic-field gradient~\cite{Medley2011spin,Jotzu2015creating} or optical Stern-Gerlach effect~\cite{Taie2010Realization}.
In frequency space, the induced global spin current $\tilde{\J}_{s}(\omega)$ at frequency $\omega$ is given by  $\tilde{\J}_{s}(\omega)=V\sigma_s(\omega)\tilde{\bm{f}}_{s}(\omega)$, where $\sigma_s(\omega)$ is the optical spin conductivity and  $\tilde{\bm{f}}_{s}(\omega)$ is the Fourier transform of $\bm{f}_{s}(t)$.
The Kubo formula provides the expression of $\sigma_s(\omega)$ as~\cite{sekino2022optical}\footnote{
In Ref.~\cite{sekino2022optical}, a perturbation given by spin-dependent scalar potentials $\delta V_{\up/\down}(\r,t)=\mp \bm{f}_{s}(t)\cdot\r$ is considered.
While there is no problem for trapped systems, formalism with this perturbation sometimes leads to incorrect results in the case of Bose systems with periodic boundary conditions.
Indeed, the contribution of condensates with zero momentum to $n/m$ in Eq.~\eqref{eq:sigma} is lost.
This problem can be avoided by expressing $\bm{f}_s(t)$ in terms of spin-dependent vector potentials in a similar way as for that of an electric field in the case of charge transport~\cite{mahan2000many,schrieffer1964theory}}
\begin{align}\label{eq:sigma}
\sigma_s(\omega)&=\frac{i}{\omega^+}\left(\frac{n}{m}+\chi_{ss}(\omega)\right),
\end{align}
where $\omega^+=\omega+i0^+$,
\begin{align}\label{eq:chi}
\chi_{ss}(\omega)=\frac{-i}{V}\int_0^\infty dt\,e^{i\omega^+ t}\<[\hat{J}_{s,x}(t),\hat{J}_{s,x}(0)]\>
\end{align}
is the spin-current response function with the spin current operator $\hat{\J}_{s}=\sum_\k\frac{\k}{m}(a_{\k,\up}^\+a_{\k,\up}-a_{\k,\down}^\+a_{\k,\down})$, and $\<\cdots\>$ denotes the thermal average at temperature $T$.
In experiments, $\sigma_s(\omega)$ can be extracted by monitoring the spin-resolved center of mass motion (see Ref.~\cite{sekino2022optical} for details).

In this paper, we focus on the real part of $\sigma_s(\omega)$ related to dissipation.
Note that $\Im\,\sigma_s(\omega)$ can be reconstructed from $\Re\,\sigma_s(\omega)$ by using the Kramers-Kronig relation.
From Eq.~\eqref{eq:sigma}, the real part has the general form of
\begin{align}\label{eq:Re_sigma}
\Re\,\sigma_s(\omega)=D_s^\mathrm{D}\delta(\omega)+\sigma_s^\mathrm{reg}(\omega),
\end{align}
where the spin Drude weight $D_s^\mathrm{D}$ and the regular part $\sigma_s^\mathrm{reg}(\omega)$ at finite frequencies are given by
\begin{align}\label{eq:D_s0}
D_s^\mathrm{D}&=D_0+\pi\Re\,\chi_{ss}(0),\\
\label{eq:sigma^reg}
\sigma_s^\mathrm{reg}(\omega)&=-\Im\,\frac{\chi_{ss}(\omega)}{\omega},
\end{align}
respectively.
The total spectral weight $D_0$ characterizes the following $f$-sum rule, which is an exact relation for the integral of $\Re\,\sigma_s(\omega)$ over $\omega$~\cite{enss2013shear,sekino2022optical}:
\begin{align}\label{eq:f-sum}
\int_{-\infty}^{\infty}\!\!d\omega\,\Re\,\sigma_s(\omega)=D_0\equiv\frac{\pi n}{m}.
\end{align}

To evaluate $\Re\,\sigma_s(\omega)$, we compute $\chi_{ss}(\omega)$ in Eq.~\eqref{eq:chi} within the Bogoliubov theory.
Using Eqs.~\eqref{eq:b} and \eqref{eq:H_2}, we can express $\hat{\J}_{s}(t)=e^{iHt}\hat{\J}_{s}e^{-iHt}$ in terms of $b_{\k,\alpha}$.
The operator has two contributions $\hat{\J}_{s}(t)=\hat{\J}_+(t)+\hat{\J}_-(t)$ given by
\begin{subequations}\label{eq:J_pm}
\begin{align}
\hat{\J}_+(t)
&=\sum_\k\frac{\k}{m}A_k^+(e^{iE_+(\epsilon_k)t}b_{\k,d}^\+b_{-\k,s}^\++\mathrm{h.c.}),\\
\hat{\J}_-(t)
&=\sum_\k\frac{\k}{m}A_k^-(e^{iE_-(\epsilon_k)t}b_{\k,d}^\+b_{\k,s}+\mathrm{h.c.}),
\end{align}
\end{subequations}
where $A_k^+=v_{k,d}u_{k,s}-u_{k,d}v_{k,s}$, $A_k^-=u_{k,d}u_{k,s}-v_{k,d}v_{k,s}$, and
\begin{align}
E_\pm(\epsilon_k)=E_d(\epsilon_k)\pm E_s(\epsilon_k)
\end{align}
determines the the time evolution of $\hat{\J}_\pm(t)$.
While $\hat{\J}_+(t)$ involves creation or annihilation of a pair of quasiparticles in density and spin modes, $\hat{\J}_-(t)$ involves transition from the spin (density) to density (spin) modes.

Here, we point out behaviors of $E_\pm(\epsilon)$ as functions of $\epsilon$.
We can easily see that $E_+(\epsilon)$ and $|E_-(\epsilon)|$ are monotonically increasing functions of $\epsilon$.
While $E_+(\epsilon)$ takes values from $0$ to $\infty$, $E_-(\epsilon)$ does from $0$ to $g_{\up\down}n$ with increasing $\epsilon>0$ and its sign depends on whether the intercomponent interaction $g_{\up\down}$ is repulsive or attractive.

Substituting Eqs.~\eqref{eq:J_pm} into Eq.~\eqref{eq:chi}, we can evaluate the current response function.
The cross terms $\<[\hat{J}_{\pm,x}(t),\hat{J}_{\mp,x}(0)]\>$ vanish and thus $\chi_{ss}(\omega)$ is found to be
\begin{align}\label{eq:chi1}
\chi_{ss}(\omega)
&=\frac{1}{V}\sum_{\nu=\pm}\sum_{\k}\frac{k^2}{3m^2}B_\nu(\epsilon_k)F_\nu(\epsilon_k)\cr
&\quad\times\left(\frac{1}{\omega^+-E_\nu(\epsilon_k)}-\frac{1}{\omega^++E_\nu(\epsilon_k)}\right).
\end{align}
Here $B_\pm(\epsilon_k)=(A_k^\pm)^2
=\frac{[E_\mp(\epsilon_k)]^2}{4E_d(\epsilon_k)E_s(\epsilon_k)}
$,
$F_+(\epsilon_k)=1+f_\mathrm{B}(E_{k,d})+f_\mathrm{B}(E_{k,s})$, 
$F_-(\epsilon_k)=f_\mathrm{B}(E_{k,s})-f_\mathrm{B}(E_{k,d})$,
where $f_\mathrm{B}(E)=1/(e^{E/T}-1)$ is the Bose distribution function.
Note that because of rotational invariance of the thermal state we replaced $k_x^2\to k^2/3$ in Eq.~\eqref{eq:chi1}.
From the distribution functions in $F_\pm(\epsilon)$, we see that the $\nu=-$ terms in Eq.~\eqref{eq:chi1} arise from thermally excited quasiparticles and vanish at zero temperature, while the $\nu=+$ ones survive even at $T=0$.
By substituting Eq.~\eqref{eq:chi1} into Eq.~\eqref{eq:sigma}, one can confirm that $\sigma_s(\omega)$ obtained in the Bogoliubov theory satisfies the $f$-sum rule in Eq.~\eqref{eq:f-sum}.

\section{\label{sec:regular}Regular part}
This section is devoted to analyzing the regular part of the optical spin conductivity.
Within the Bogoliubov theory, $\sigma_s^\mathrm{reg}(\omega)$ can be analytically computed.
Since $\sigma_s^\mathrm{reg}(\omega)$ is an even function of $\omega$ by definition, we can focus on the case of $\omega>0$ without loss of generality.
For $g_{\up\down}>0$, substituting Eq.~\eqref{eq:chi1} into Eq.~\eqref{eq:sigma^reg} yields
\begin{align}\label{eq:sigma^reg1}
\sigma_s^\mathrm{reg}(\omega)&=\frac{\sqrt{2m}}{3\pi}\sum_{\nu=\pm}\int_0^\infty\!\!d\epsilon\,\frac{\epsilon^{3/2}B_\nu(\epsilon)F_\nu(\epsilon)}{E_\nu(\epsilon)}\cr
&\qquad\qquad\qquad\times\delta(\omega-E_\nu(\epsilon)).
\end{align}
The delta function shows that $\sigma_s^\mathrm{reg}(\omega)$ is sensitive to the sum and difference of the quasiparticle energies.
As mentioned above, $E_\nu(\epsilon)$ are monotonic functions of $\epsilon$.
The equation $E_+(\epsilon)=\omega$ has a single solution $\epsilon=\epsilon^+_\omega$ for any $\omega>0$, while the equation $E_-(\epsilon)=\omega$ does $\epsilon=\epsilon^-_\omega$ only if $0<\omega<g_{\up\down}n$.
These solutions are given by
\begin{align}
\label{eq:16}
\epsilon^\pm_\omega=\frac{\omega^2}{2\left(gn\pm\sqrt{4g_dg_sn^2+\omega ^2}\right)}.
\end{align}
By using Eq.~\eqref{eq:16}, the analytic form of $\sigma_s^\mathrm{reg}(\omega)$ can be obtained.
The regular part for $g_{\up\down}<0$ is also calculated in a similar way.
Finally, $\sigma_s^\mathrm{reg}(\omega)$ is found to be
\begin{align}\label{eq:sigma^reg2}
\sigma_s^\mathrm{reg}(\omega)&=\sigma_s^+(\omega)+\sigma_s^-(\omega)\theta(|g_{\up\down}|n-\omega),
\end{align}
where
\begin{align}\label{eq:sigma^+-}
\sigma_s^\nu(\omega)=\frac{\sqrt{2m\,(\epsilon_\omega^\nu)^{3}}}{3\pi\omega}\frac{B_\nu(\epsilon_\omega^\nu)\left|F_\nu(\epsilon_\omega^\nu)\right|}{G_\nu(\epsilon_\omega^\nu)}
\end{align}
is the contribution from the $\nu=\pm$ process with $G_\pm(\epsilon)=\left|\frac{\epsilon+g_dn}{E_d(\epsilon)}\pm\frac{\epsilon+g_sn}{E_s(\epsilon)}\right|$.
The Heaviside step function $\theta(|g_{\up\down}|n-\omega)$ in Eq.~\eqref{eq:sigma^reg2} arises from the upper limit of $|E_-(\epsilon)|$, so that $\sigma_s^-(\omega)$ contributes to the spin conductivity in a low frequency regime.
The result of Eq.~\eqref{eq:sigma^reg2} is independent of the sign of the intercomponent coupling $g_{\up\down}\propto a_{\up\down}$.
Such insensitivity to the sign is also pointed out in the context of the Andreev-Bashkin effect within the Bogoliubov theory~\cite{fil2005nondissipative}.
For this reason, we consider only the case of $0<a_{\up\down}\leq a$ in Figs.~\ref{fig:sigma_T=0}--\ref{fig:eta} below without loss of generality.

\subsection{Zero-temperature case}
\begin{figure}[tb]
\includegraphics[width=0.98\hsize]{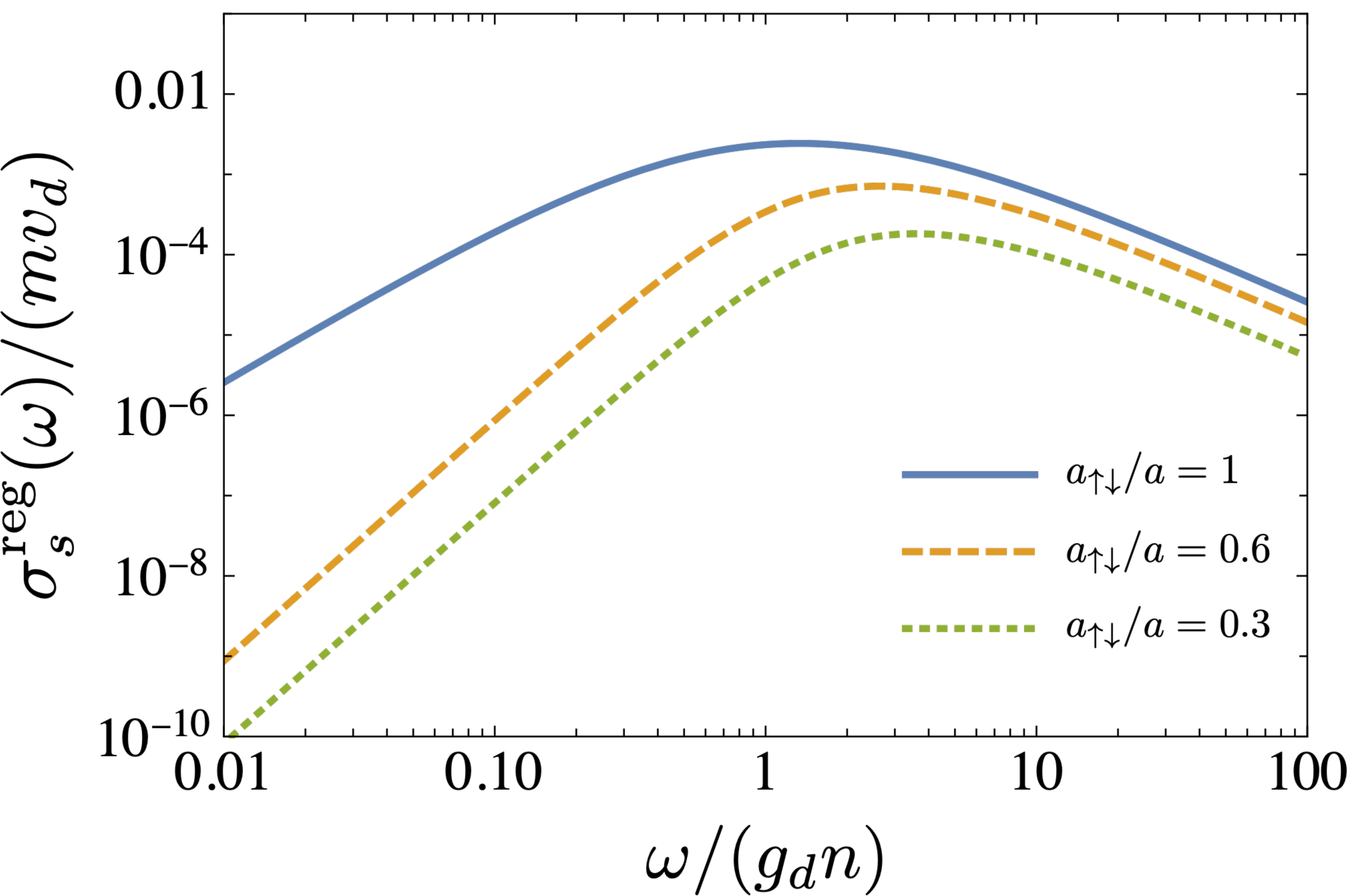}
\caption{\label{fig:sigma_T=0}Spectrum of the regular part of the optical spin conductivity $\sigma_s^\mathrm{reg}(\omega)=\sigma_s^+(\omega)$ at $T=0$.
The low-frequency power law at the transition point ($a_{\up\down}=a$) is different from those inside the miscible phase ($0<a_{\up\down}<a$) [see Eqs.~\eqref{eq:sigma_small_T=0}], while the power law at high frequency is identical [Eq.\eqref{eq:sigma_large}].}
\end{figure}

First, we focus on $\sigma_s^\mathrm{reg}(\omega)$ at zero temperature.
In this case, $\sigma_s^-(\omega)$ in Eq.~\eqref{eq:sigma^reg2} vanishes because it results from thermally excited quasiparticles, leading to
\begin{align}\label{eq:sigma^reg_T=0}
\sigma_s^\mathrm{reg}(\omega)=\frac{\sqrt{2m\,(\epsilon_\omega^+)^{3}}}{3\pi\omega}\frac{B_+(\epsilon_\omega^+)}{G_+(\epsilon_\omega^+)}.
\end{align}
Figure~\ref{fig:sigma_T=0} shows the spectra of $\sigma_s^\mathrm{reg}(\omega)$ for various interaction strengths, where $v_\alpha=\sqrt{g_\alpha n/m}$ is the sound velocity of the $\alpha$ mode.
The spectrum for $a_{\up\down}/a=1$ corresponds to the result at the transition point, while those at $a_{\up\down}/a=0.3,\,0.6$ are for the Bose mixtures inside the miscible phase.
As the ratio $a_{\up\down}/a$ increases, $\sigma_s^\mathrm{reg}(\omega)$ is enhanced in the whole frequency regime.
This is caused by the fact that $\sigma_s^\mathrm{reg}(\omega)$ reflects scatterings between different spin components.
With $a_{\up\down}/a>0$ increasing the effect of the scatterings becomes stronger, so that $\sigma_s^\mathrm{reg}(\omega)$ is enhanced.
One can also see that $\sigma_s^\mathrm{reg}(\omega)$ at $T=0$ has a peak around $\omega\sim g_dn$, where transition of quasiparticle excitations between phonon-like to free-particle-like regimes occurs.

Slopes in the low frequency regime in Fig.~\ref{fig:sigma_T=0} imply that $\sigma_s^\mathrm{reg}(\omega\to+0)$ for $|a_{\up\down}|<a$ and $a_{\up\down}=\pm a$ obeys different power laws.
Indeed, expanding Eq.~\eqref{eq:sigma^reg_T=0} in small $\omega$ yields
\begin{align}\label{eq:sigma_small_T=0}
\sigma_s^\mathrm{reg}(\omega\to+0)
=
\begin{dcases}
\frac{C_1(\gamma)}{m^2v_d^5}\omega^3&(|a_{\up\down}|<a),\\
\frac{\omega^2}{12\pi m\tilde{v}^3}&(a_{\up\down}=\pm a),
\end{dcases}
\end{align}
where $\gamma=g_s/g_d=(a-a_{\up\down})/(a+a_{\up\down})$, $\tilde{v}=\sqrt{gn/m}$, and
\begin{align}
C_1(\gamma)&=\frac{1}{24\pi}\frac{(1-\sqrt{\gamma})^2}{\sqrt{\gamma}(1+\sqrt{\gamma})^5}.
\end{align}
This change of the power law at $a_{\up\down}=\pm a$ is related to excitation properties of quasiparticles specific to the phase transition points.
As mentioned previously, $\sigma_s^\mathrm{reg}(\omega)$ is sensitive to the quasiparticle spectra.
In the case of the mixture inside the miscible phase ($|a_{\up\down}|<a$), both density and spin modes show linear dispersions.
On the other hand, at $a_{\up\down}=\pm a$, one of the modes has a quadratic dispersion $E_\alpha(\epsilon_k)=\epsilon_k$, while the other still shows the linear behavior.
Equation~\eqref{eq:sigma_small_T=0} implies that the precursors of the phase transitions are captured by the change of the low-frequency behavior of $\sigma_s^\mathrm{reg}(\omega)$~\footnote{More specifically, the different power laws result from the fact that two limits $\omega\to+0$ and $g_{\up\down}\to\pm g$ for $B_+(\epsilon_\omega^+)$ do not commute.}.

At high frequency, the dispersion relations of quasiparticles contributing to $\sigma_s^\mathrm{reg}(\omega\to\infty)$ become quadratic in momenta [$E_\alpha(\epsilon_k)\simeq\epsilon_k$], which lead to a power-law tail of the optical spin conductivity.
By expanding Eq.~\eqref{eq:sigma^reg_T=0} in large $\omega$, we obtain
\begin{align}\label{eq:sigma_large}
\sigma_s^\mathrm{reg}(\omega\to\infty)=\frac{\sqrt{m}(g_{\up\down}n)^2}{12\pi \omega^{3/2}}.
\end{align}
This frequency dependence $\sim\omega^{-3/2}$ is the same as those of three-dimensional Fermi gases in both normal and superfluid phases~\cite{enss2012quantum,hofmann2011current,sekino2022optical}, a spinor BEC in the polar phase~\cite{sekino2022optical}, and a one-dimensional Fermi superfluid with $p$-wave attraction~\cite{tajima2022optical}.

\subsection{Finite-temperature case}
\begin{figure}[tb]
\includegraphics[width=0.98\hsize]{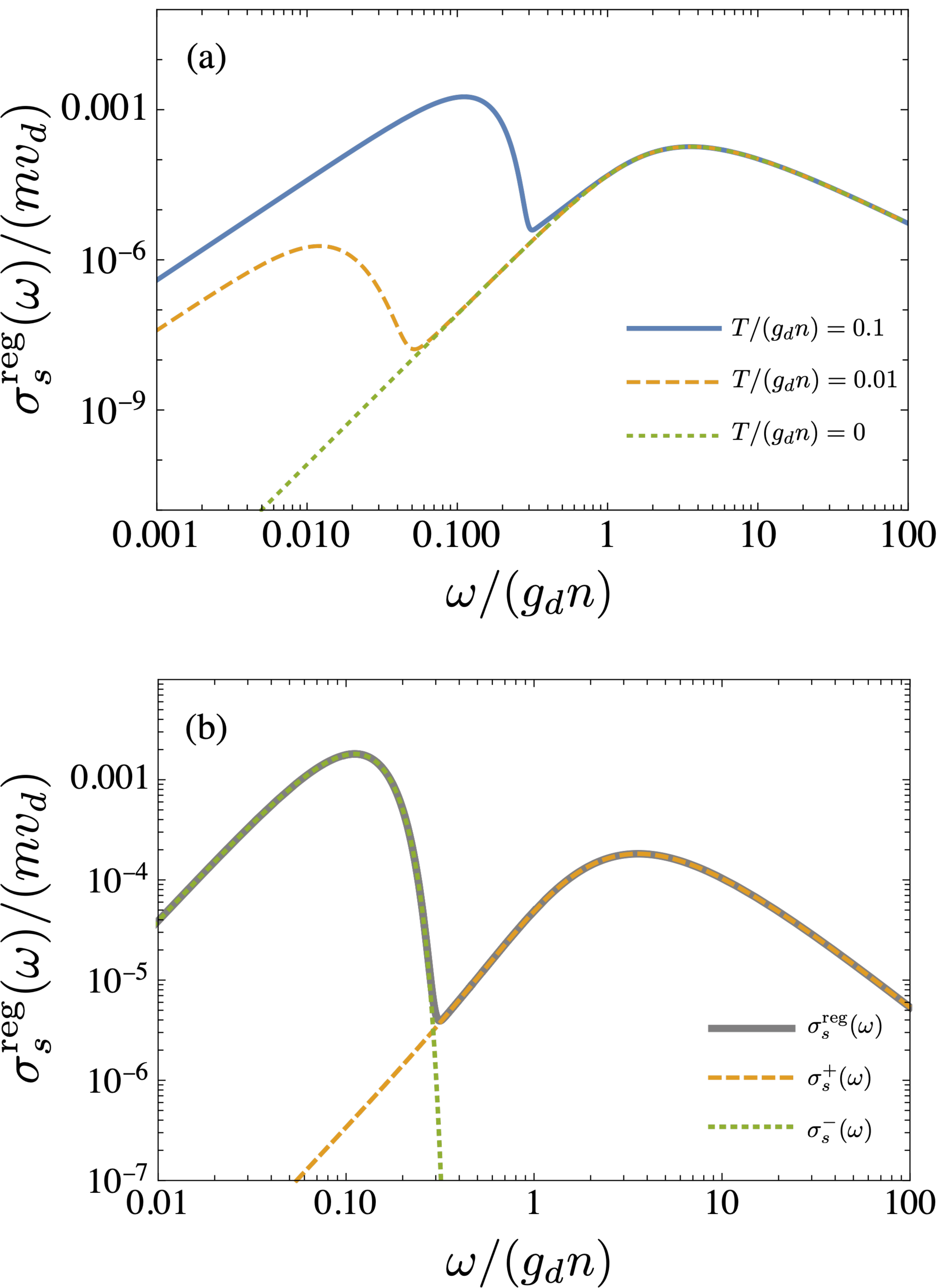}
\caption{\label{fig:sigma_T>0}Regular part of the optical spin conductivity $\sigma_s^\mathrm{reg}(\omega)$ at various $T$ with $a_{\up\down}/a=0.3$ fixed.
(a) Temperature dependence of $\sigma_s^\mathrm{reg}(\omega)$. 
At finite temperature, an additional peak appears around $\omega\sim T$.
(b) Contributions of $\sigma_s^\pm(\omega)$ to $\sigma_s^\mathrm{reg}(\omega)$ at $T/(g_dn)=0.1$ and $a_{\up\down}/a=0.3$.}
\end{figure}
We next discuss the regular part at a finite but sufficiently low temperature.
In this case, $\sigma_s^-(\omega)$ in Eq.~\eqref{eq:sigma^reg2} contributes to the spin conductivity in a low-frequency regime.
First, we start with analysis of asymptotic behaviors at low frequencies.
At $T>0$, the appearance of the temperature scale modifies the power law at small $\omega$ in a similar way as for the momentum distribution at small momentum~\cite{pitaevskii2003}.
Expanding Eq.~\eqref{eq:sigma^reg2}, we obtain
\begin{align}\label{eq:sigma_small_T>0}
\sigma_s^\mathrm{reg}(\omega\to+0)=
\begin{dcases}
\frac{C_2(\gamma)T}{m^2v_d^5}\omega^2&(|a_{\up\down}|<a),\\
\frac{T}{3\pi \tilde{v}}&(a_{\up\down}=\pm a),
\end{dcases}
\end{align}
where
\begin{align}
C_2(\gamma)&=\frac{1}{24\pi}\frac{(1+\sqrt{\gamma})^5+|1-\sqrt{\gamma}|^5}{\,\gamma|1-\gamma|^3}.
\end{align}
Compared with Eq.~\eqref{eq:sigma_small_T=0} at $T=0$, $\sigma_s^\mathrm{reg}(\omega\to+0)$ decays more slowly for the mixture inside the miscible phase $(|a_{\up\down}|<a)$, while the spectra at the transition points $(a_{\up\down}=\pm a)$ exhibit plateaus in the low-frequency regime.
Furthermore, the asymptotic forms in Eq.~\eqref{eq:sigma_small_T>0} indicate that the magnitude of $\sigma_s^\mathrm{reg}(\omega)$ is enhanced as temperature increases.
This results from the increasing numbers of thermally excited quasiparticles relevant to the $\nu=-$ process.
On the other hand, the asymptotic form  at high frequency is identical with Eq.~\eqref{eq:sigma_large} in the zero-temperature case and not sensitive to $T$.
This is because at high frequency $\sigma_s^-(\omega)$ never contributes and the thermal effect on $\sigma_s^+(\omega)$ vanishes [$F_+(\epsilon)\simeq1$].

\begin{figure}[tb]
\includegraphics[width=0.98\hsize]{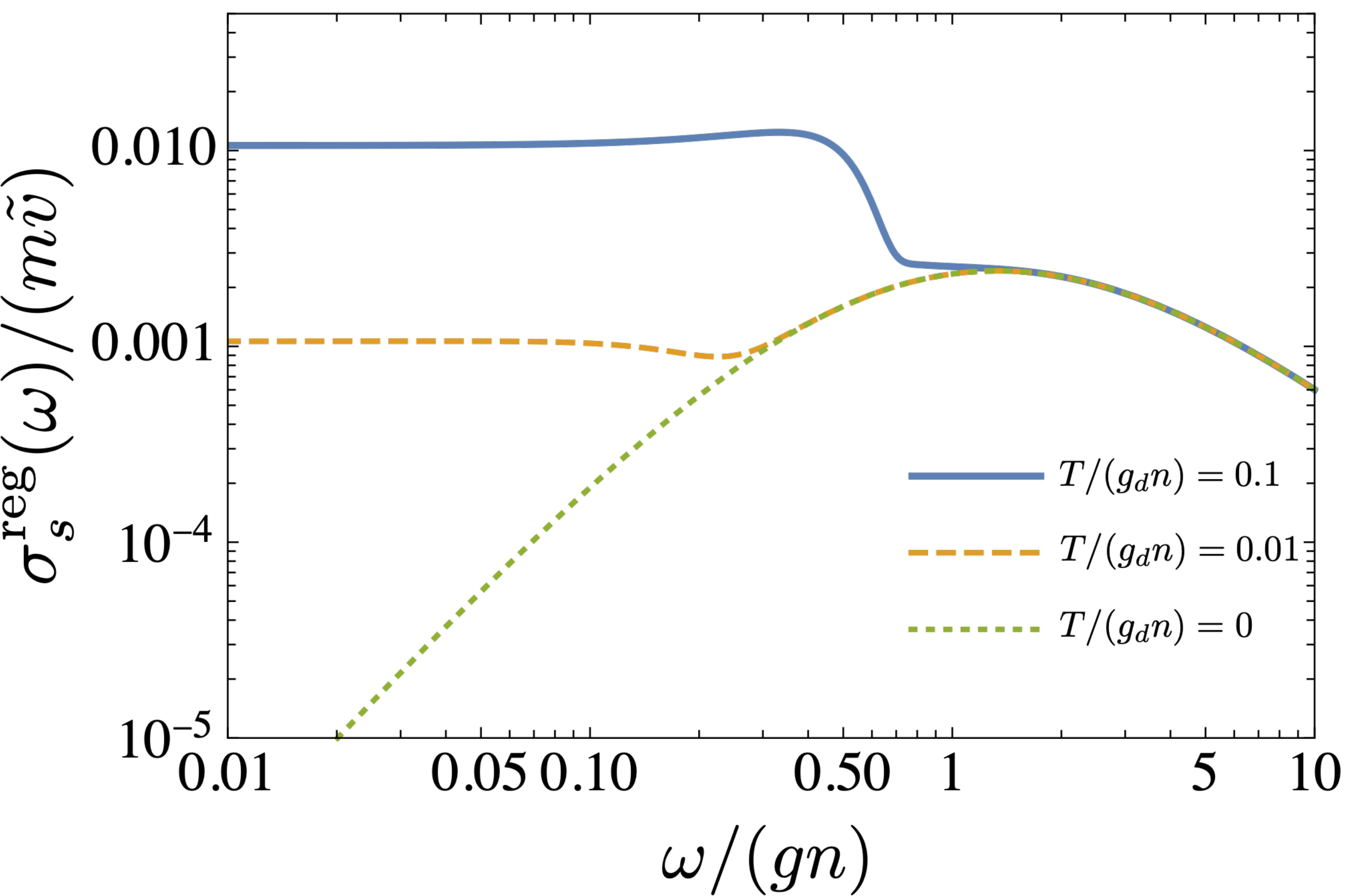}
\caption{\label{fig:sigma_transition}Spectrum of the regular part $\sigma_s^\mathrm{reg}(\omega)$ at the phase transition point $a_{\up\down}=a$.}
\end{figure}

Figure~\ref{fig:sigma_T>0} illustrates the impact of temperature on $\sigma_s^\mathrm{reg}(\omega)$ inside the miscible phase with $a_{\up\down}/a=0.3$ fixed.
In Fig.~\ref{fig:sigma_T>0}a, $\sigma_s^\mathrm{reg}(\omega)$ at $T>0$ has two peaks around $\omega\sim T$ and $\omega\sim g_dn$.
The peak in the lower frequency side arises from $\sigma_s^-(\omega)$ associated with thermally excited quasiparticles (see Fig.~\ref{fig:sigma_T>0}b).
As implied by Eq.~\eqref{eq:sigma_small_T>0}, this peak is enhanced as temperature increases.
In particular, the peak at $T/(g_dn)=0.1$ becomes higher than the other.
On the other hand, the peak in the higher frequency side arises from $\sigma_s^+(\omega)$ (see Fig.~\ref{fig:sigma_T>0}b) and is not sensitive to $T$ because distribution functions in $F_+(\epsilon_\omega^+)$ are vanishingly small for $\omega\sim g_dn\gg T$.
Figure~\ref{fig:sigma_transition} shows the temperature dependence of $\sigma_s^\mathrm{reg}(\omega)$ at the phase transition point $a_{\up\down}=a$.
As presented by Eq.~\eqref{eq:sigma_small_T>0}, $\sigma_s^\mathrm{reg}(\omega)$ at $T>0$ exhibits the plateau at low frequency, which becomes higher with increasing $T$.

In the above analysis based on the Bogoliubov theory, we focused on the low-temperature regime of a weakly interacting Bose mixture, where depletion of the condensate density is negligible.
In this case, the thermal effect on $\sigma_s^\mathrm{reg}(\omega)$ arises only through distribution functions in $F_\pm(\epsilon_\omega^\pm)$.
For a higher temperature, on the other hand, the thermal depletion becomes significant and the peak of $\sigma_s^\mathrm{reg}(\omega)$ around $\omega\sim g_dn$ would be affected by the thermal effects.

\section{\label{sec:Drude}Spin Drude weight}
In this section, we discuss the spin Drude weight $D_s^\mathrm{D}$ and its connection with the Andreev-Bashkin effect.
As mentioned previously, the $\mathbb{Z}_2$ symmetric mixture is considered as a spin superfluid~\cite{flayac2013superfluidity,fava2018observation,kim2020obvervation}.
To clarify the importance of $D_s^\mathrm{D}$ in a spin superfluid, we start with a brief review of work by Scalapino, White, and Zhang, who provided the criteria to determine whether an electron system is superconducting, metallic, or insulating~\cite{Scalapino1992superfluid,scalapino1993insulator}.
This previous work has suggested that a superconductor, metal, and insulator are distinguished by the two properties in electric transport.
The first one is the Drude weight $D^\mathrm{D}$, which characterizes the delta-function peak of optical conductivity $\sigma(\omega)$ at $\omega=0$ as the spin Drude weight $D_s^\mathrm{D}$ in Eq.~\eqref{eq:Re_sigma}.
A finite $D^\mathrm{D}$ corresponds to ballistic charge transport with diverging dc conductivity, i.e., zero resistivity.
The latter one is the superfluid weight $D^\mathrm{SF}$ proportional to the superfluid density relevant to the Meissner effect for a charged system.

At $T=0$, a superconductor has finite $D^\mathrm{D}$ 
 and $D^\mathrm{SF}$, a metal without impurities has finite $D^\mathrm{D}$ and $D^\mathrm{SF}=0$, and an insulator has $D^\mathrm{D}=D^\mathrm{SF}=0$.
Even at finite temperature, a superconductor still has $D^\mathrm{D}>0$ and $D^\mathrm{SF}>0$ and exhibits dissipationless transport as long as the corresponding order exists.
On the other hand, a metal at finite temperature or in the presence of impurities is generally expected to have $D^\mathrm{D}=0$ and become resistive with a finite dc conductivity $\sigma(\omega\to0)>0$~\footnote{Integrable systems are considered to be exceptional cases with finite Drude weights at $T>0$~\cite{bulchandani2021superdiffusion}.}.
In this way, $D^\mathrm{D}$ and $D^\mathrm{SF}$ distinguish a superconductor from a metal and insulator in terms of electric transport properties.
Table~\ref{table:1} summarizes classification of ground states by the Drude and superfluid weights.

Similarly, the spin Drude weight $D_s^\mathrm{D}$ and spin superfluid weight $D_s^\mathrm{SF}$ are essential to identify a quantum many-body state as a spin superfluid.
In analogy with charge transport, the spin superfluidity is characterized by both $D_s^{\mathrm{D}}>0$ and $D_s^{\mathrm{SF}}>0$, while a ground state with $D_s^{\mathrm{D}}>0$ and $D_s^{\mathrm{SF}}=0$ and that with $D_s^{\mathrm{D}}=D_s^{\mathrm{SF}}=0$ may be referred to as spin metal and spin insulator, respectively, which are not discussed mainly in this paper.
In the next subsection, we evaluate $D_s^\mathrm{D}$ within the Bogoliubov theory.
On the other hand, the discussion on $D_s^\mathrm{SF}$ for the binary mixture is presented in Appendix~\ref{appendix:superfluid_weights}.
\begin{table}
\caption{Summary of Drude and superfluid weights in charge (upper) and spin (lower) transport at $T=0$}.
\begin{ruledtabular}
\begin{tabular}{ccc}
&---Charge transport---&
\\
Superconductor & Metal & Insulator \\
\hline
$D^{\mathrm{D}}>0$, $D^{\mathrm{SF}}>0$  & $D^{\mathrm{D}}>0$, $D^{\mathrm{SF}}=0$ & $D^{\mathrm{D}}=D^{\mathrm{SF}}=0$ \\
\hline
\hline\\
&---Spin transport---& 
\\
Spin superfluid & Spin metal & Spin insulator\\
\hline
$D_s^{\mathrm{D}}>0$, $D_s^{\mathrm{SF}}>0$  & $D_s^{\mathrm{D}}>0$, $D_s^{\mathrm{SF}}=0$ & $D_s^{\mathrm{D}}=D_s^{\mathrm{SF}}=0$ \\
\end{tabular}
\end{ruledtabular}
\label{table:1}
\end{table}

\subsection{Computation within the Bogoliubov theory}
We first rewrite Eq.~\eqref{eq:D_s0} by using the Kramers-Kronig relation for $\chi_{ss}(\omega)$ and $\sigma_s^\mathrm{reg}(\omega)=\sigma_s^\mathrm{reg}(-\omega)$:
\begin{align}\label{eq:D_s^D}
D_s^\mathrm{D}
&=D_0-D_s^\mathrm{reg},\\
\label{eq:D_s^reg}
D_s^\mathrm{reg}&=2\int_0^\infty\!\!d\omega\,\sigma_s^\mathrm{reg}(\omega).
\end{align}
These equations mean that $D_s^\mathrm{D}$ is reduced from the total spectral weight $D_0$ due to the weight $D_s^\mathrm{reg}$ originating from the regular part.
To compute $D_s^\mathrm{reg}$, we substitute Eq.~\eqref{eq:sigma^reg1} into Eq.~\eqref{eq:D_s^reg}, leading to
\begin{align}\label{eq:eta}
\frac{D_s^\mathrm{reg}}{D_0}
=\frac{2\sqrt{2m^3}}{3\pi^2n}\sum_{\nu=\pm}\int_0^\infty\!\!d\epsilon\,\frac{\epsilon^{3/2}B_\nu(\epsilon)F_\nu(\epsilon)}{E_\nu(\epsilon)},
\end{align}
which holds in the whole range of $0<|a_{\up\down}|\leq a$.
For convenience, we define $a_\alpha$ by $g_\alpha=\frac{4\pi a_\alpha}{m}$ and a scaled weight
$\tilde{D}_s^\mathrm{reg}=D_s^\mathrm{reg}/(D_0\sqrt{na_d^3})$, which is a dimensionless function of $a_{\up\down}/a$ and $T/g_dn$.
\begin{figure}[tb]
\includegraphics[width=0.98\hsize]{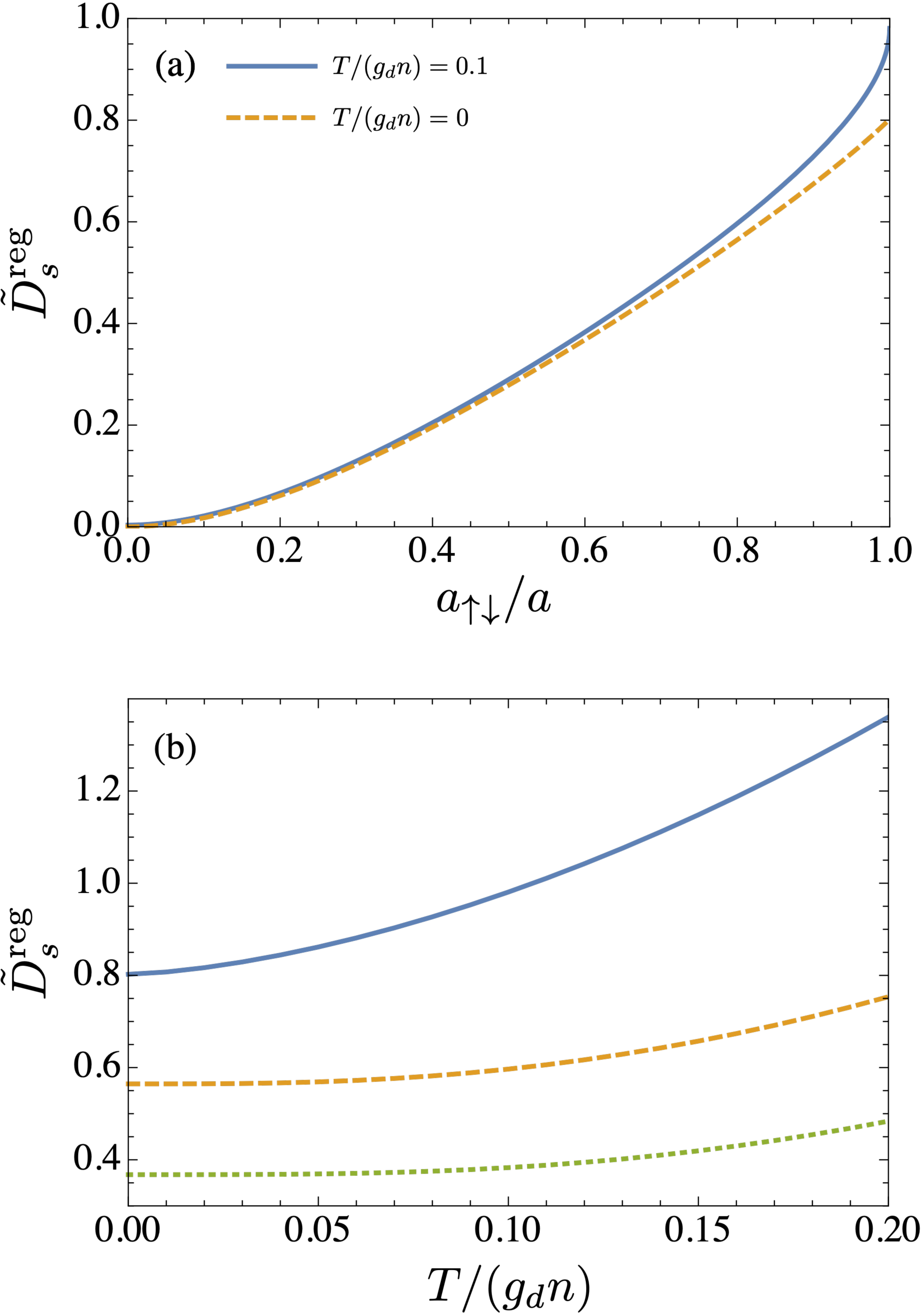}
\caption{\label{fig:eta}Scaled weight $\tilde{D}_s^\mathrm{reg}=D_s^\mathrm{reg}/(D_0\sqrt{na_d^3})$ of the regular part.
(a) Dependence on $a_{\up\down}/a$ with $T$ fixed.
$\tilde{D}_s^\mathrm{reg}$ takes its maximum value at the transition point $a_{\up\down}/a=1$.
(b) Dependence on $T$ for $a_{\up\down}/a=1$ (blue solid line), $a_{\up\down}/a=0.8$ (yellow dashed line), and $a_{\up\down}/a=0.6$ (green dotted line).}
\end{figure}

Figure~\ref{fig:eta} shows the scaled weight $\tilde{D}_s^\mathrm{reg}$ of the regular part for $0<a_{\up\down}\leq a$ at low temperature.
Figure~\ref{fig:eta}a indicates how $\tilde{D}_s^\mathrm{reg}$ depends on the dimensionless strength of the intercomponent interaction $a_{\up\down}/a$ with $T$ fixed.
At $T=0$, we can analytically perform the integration in Eq.~\eqref{eq:eta} and find the following expression:
\begin{align}\label{eq:eta_T=0}
\tilde{D}_s^\mathrm{reg}=\tilde{D}_{s,0}^\mathrm{reg}(\gamma)\equiv\frac{64}{45\sqrt{\pi}}\frac{\left(1-\sqrt{\gamma}\right)^2 \left(\gamma+3\sqrt{\gamma}+1\right)}{\sqrt{\gamma}+1}
\end{align}
with $\gamma=(a-a_{\up\down})/(a+a_{\up\down})$.
As $a_{\up\down}/a$ increases, $\tilde{D}_s^\mathrm{reg}$ grows and becomes maximum at the transition point $a_{\up\down}/a=1$, which is consistent with enhancement of the spectra in Fig.~\ref{fig:sigma_T=0}.
With increasing $a_{\up\down}/a$ the effect of scatterings between different spin components becomes stronger, leading to the growth of $\tilde{D}_s^\mathrm{reg}$.
At $a_{\up\down}/a=0$, where the intercomponent interaction vanishes, $\tilde{D}_s^\mathrm{reg}$ becomes zero.
This is because in this case the global spin current $\hat{\J}_{s}(t)$ is a conserved quantity, leading to a trivial form of the conductivity $\Re\,\sigma_s(\omega)=D_0\delta(\omega)$.

Figure~\ref{fig:eta}b shows the temperature dependence of $\tilde{D}_s^\mathrm{reg}$ with $a_{\up\down}/a$ fixed.
The weight of the regular part increases with increasing temperature as expected from the enhancement of the spectrum in Fig.~\ref{fig:sigma_T>0}a.
Inside the miscible phase $0<a_{\up\down}<a$, the temperature dependence of $\tilde{D}_s^\mathrm{reg}$ is very weak at low temperature as shown by yellow dashed and green dotted lines in Fig.~\ref{fig:eta}b.
In fact, by expanding Eq.~\eqref{eq:eta} with respect to $T$, we obtain
\begin{align}
\tilde{D}_s^\mathrm{reg}\simeq\tilde{D}_{s,0}^\mathrm{reg}(\gamma)+C_3(\gamma)\left(\frac{T}{g_dn}\right)^4,
\end{align}
with 
\begin{align}
C_3(\gamma)=\frac{4 \pi ^{7/2} \left(\gamma ^3+\gamma ^{5/2}+4 \gamma ^2+4 \gamma ^{3/2}+4 \gamma +\sqrt{\gamma }+1\right) }{45 \left(\sqrt{\gamma }+1\right) \gamma ^{5/2}}.
\end{align}
On the other hand, at $a_{\up\down}=a$ (the blue solid line in Fig.~\ref{fig:eta}b), $\tilde{D}_s^\mathrm{reg}$ is more sensitive to $T$ as
\begin{align}\label{eq:eta_T}
\tilde{D}_s^\mathrm{reg}\simeq\tilde{D}_{s,0}^\mathrm{reg}(0)+\frac{4 \sqrt{2} \zeta \left(\frac{3}{2}\right)}{3}\left(\frac{T}{g_d n}\right)^{3/2}.
\end{align}

We now discuss the value of the spin Drude weight $D_s^\mathrm{D}$.
The above evaluations show $\tilde{D}_s^\mathrm{reg}\sim1$ at sufficiently low temperature ($T\ll g_dn$), leading to $D_s^\mathrm{reg}\sim D_0\sqrt{na^3}\ll D_0$ for the weakly interacting mixture.
Therefore, $D_s^\mathrm{D}$ in Eq.~\eqref{eq:D_s^D} is found to be finite $D_s^\mathrm{D}>0$, indicating that the binary mixture of BECs exhibits a zero spin resistivity resulting from spin superfluidity.
It is worth emphasizing the difference of $D_s^\mathrm{D}$ from the Drude weight $D_n^\mathrm{D}$ in optical mass conductivity.
Because of the Kohn's theorem~\cite{kohn1961cyclotron}, the optical mass conductivity gives $D_n^\mathrm{D}=D_0$ for any scattering lengths and temperature.
In contrast, Eq.~\eqref{eq:D_s^D} indicates that $D_s^\mathrm{D}$ is reduced from $D_0$ because the spectral weight is transferred to the regular part $\sigma_s^\mathrm{reg}(\omega)$ at finite frequency due to the interactions.
As $|a_{\up\down}|$ or $T$ increase with $\sqrt{na^3}\ll1$ fixed, the excitations at finite momenta contributing to $D_s^\mathrm{reg}$ are enhanced and thus $D_s^\mathrm{D}$ decreases.
As in the case of superconductor with finite $D^\mathrm{D}$~\cite{scalapino1993insulator}, $D_s^\mathrm{D}$ is however expected to survive at higher temperature as long as the mixture is in the miscible phase of BECs.
On the other hand, $D_s^\mathrm{D}$ is expected to vanish with finite spin resistivity above the transition temperature.
In this way, one can see that the optical spin conductivity including $D_s^\mathrm{D}$ are useful probes for spin superfluid properties of homogeneous mixtures.

\subsection{\label{sec:drag}Connection with the Andreev-Bashkin effect}
Here, we discuss the relation of the spin Drude weight to the spin superfluid weight $D_s^\mathrm{SF}$ and the Andreev-Bashkin effect.
We start by considering difference of definitions between $D_s^\mathrm{D}$ and $D_s^\mathrm{SF}$.
As shown in Appendix~\ref{appendix:superfluid_weights}, 
both weights are written as limiting behaviors of $D_s(\q_\mathrm{T},\q_\mathrm{L},\omega)=D_0+\pi\Re\,\chi_{ss}(\q,\omega)$, 
where $\chi_{ss}(\q,\omega)$ is a momentum-resolved current response function in Eq.~\eqref{eq:chi_qomega} and $\q_\mathrm{T}$ and $\q_\mathrm{L}$ are the components of momentum $\q$ perpendicular and parallel to the current, respectively.
By taking limits $\q_\mathrm{T}\to\0$, $\q_\mathrm{L}\to\0$, and $\omega\to0$ in different order, we obtain $D_s^\mathrm{D}=D_s(\0,\0,\omega\to0)$ and $D_s^\mathrm{SF}=D_s(\q_\mathrm{T}\to\0,\0,0)$.

The spin superfluid weight $D_s^\mathrm{SF}$ is related to the Andreev-Bashkin drag density $\rho_{\up\down}$ characterizing how sensitive the mass current of one component is to the superfluid velocity of the other.
As shown in Appendix~\ref{appendix:superfluid_weights}, 
one can express $D_s^\mathrm{SF}$ in terms of $\rho_{\up\down}$ and the normal fluid density $\rho^\mathrm{NF}$ as
\begin{align}\label{eq:D_s^SF_rho}
    D_s^\mathrm{SF}=D_0-\frac{\pi}{m^2}\left(4\rho_{\up\down}+\rho^\mathrm{NF}\right).
\end{align}
At $T=0$ with $\rho^\mathrm{NF}=0$, $D_s^\mathrm{SF}$ is reduced from $D_0$ due to the drag differently from its mass counterpart $D_n^\mathrm{SF}=D_0$.
At finite temperature, the appearance of the normal fluid also decreases the spin superfluid weight.
The drag and normal fluid densities have been computed in Refs.~\cite{fil2005nondissipative,Ota2020thermodynamics,romito2021linear}.

Applying the Bogoliubov theory to $\chi_{ss}(\q,\omega)$, we find $D_s^\mathrm{D}=D_s^\mathrm{SF}$ at $T\geq0$.
The detailed analyses are presented in Appendix~\ref{appendix:current-response}.
We emphasize that this equivalence between Drude and superfluid weights is not self-evident.
In general, it is not guaranteed that the two limits $\q_\mathrm{T}\to0$ and $\omega\to0$ are commutable.
Indeed, there are several examples of gapless systems where the Drude and superfluid weights for mass take different values.
A homogeneous ideal Fermi gas has $D_n^\mathrm{D}=D_0$ and $D_n^\mathrm{SF}=0$, exhibiting metallic features.
For the two-component BEC focused on in this paper, Kohn's theorem provides $D_n^\mathrm{D}=D_0$ for any $T$, while $D_n^\mathrm{SF}<D_0$ is sensitive to the normal-fluid density at finite temperature (see Appendixes~\ref{appendix:superfluid_weights} and \ref{appendix:current-response} for details).
On the other hand, there is a rigorous proof that the Drude and superfluid weights are identical with each other for a gapped systems at $T=0$~\cite{scalapino1993insulator}.

Finally, we discuss the connection of the optical spin conductivity to the Andreev-Bashkin effect.
At $T=0$ with $\rho^\mathrm{NF}=0$, $D_s^\mathrm{D}=D_s^\mathrm{SF}$ combined with Eqs.~\eqref{eq:D_s^D} and \eqref{eq:D_s^SF_rho} provides
\begin{align}\label{eq:D_s^reg-rho_updown_T=0}
D_s^\mathrm{reg}=\frac{4\pi}{m^2}\rho_{\up\down}.
\end{align}
This result states that the spectral weight of $\Re\,\sigma_s(\omega)$ at finite $\omega$ corresponds to the drag density $\rho_{\up\down}$.
Therefore, the optical spin conductivity could be a probe to observe the Andreev-Bashkin effect, which has yet to
be confirmed in experiments.

The relation $D_s^\mathrm{D}=D_s^\mathrm{SF}$ proposes that measurement of the optical spin conductivity is useful to detect the Andreev-Bashkin effect even at finite temperature.
At $T>0$, the drag density is rewritten as
\begin{align}\label{eq:rho_updown_T>0}
    \rho_{\up\down}=\frac{m^2}{4\pi}D_s^\mathrm{reg}-\frac{1}{4}\rho^\mathrm{NF}.
\end{align}
When temperature is sufficiently low ($T\ll gn$), $\rho_{\up\down}$ is still estimated by measuring $D_s^\mathrm{reg}$ because $\rho^\mathrm{NF}=\order{T^4}$ is small as in the single-component case~\cite{pitaevskii2003}.
At higher temperature, the normal fluid fraction is not negligible and $\rho_{\up\down}$ deviates from $\frac{m^2}{4\pi}D_s^\mathrm{reg}$.
Indeed, $\rho_{\up\down}$ is a decreasing functions of $T$~\cite{Ota2020thermodynamics}, while Fig.~\ref{fig:eta}b indicates $D_s^\mathrm{reg}$ as an increasing function of $T$.
Over the temperature regime where the Bogoliubov theory is applicable yet $\rho^\mathrm{NF}$ is not negligible in Eq.~\eqref{eq:rho_updown_T>0}, $\rho_{\up\down}$ is experimentally accessible by measuring both $D_s^\mathrm{reg}$ and $\rho^\mathrm{NF}$.
We expect that $\rho^\mathrm{NF}$ of the binary BEC can be experimentally determined in a similar way as for single-component bosons~\cite{Christodoulou:2021aa} and fermions~\cite{Sidorenkov:2013aa,yan2022thermography}.
It would be an interesting future work whether Eq.~\eqref{eq:rho_updown_T>0} or $D_s^\mathrm{D}=D_s^\mathrm{SF}$ is valid beyond the Bogoliubov theory.

\section{\label{sec:conclusion}Conclusion}
In this paper, we investigated the optical spin conductivity for a binary mixture of BECs in the miscible phase.
The regular part of the spin conductivity was analytically evaluated with the Bogoliubov theory in Sec.~\ref{sec:regular}.
Reflecting the two gapless modes specific to the Bose mixture, two processes $\nu=\pm$ are relevant to the spin conductivity spectrum.
At zero temperature, the regular part $\sigma_s^\mathrm{reg}(\omega)$ obeys  power laws for both high- and low-frequency regimes [Eqs.~\eqref{eq:sigma_small_T=0} and \eqref{eq:sigma_large}] and exhibits a peak in the intermediate frequency regime (see Fig.~\ref{fig:sigma_T=0}).
In particular, the power law at low frequency changes at the transition points ($a_{\up\down}=\pm a$), which results from the fact that one of the gapless modes has the parabolic dispersion.
At finite temperature, the $\nu=-$ process associated with thermal excitations of quasiparticles contributes to the regular part, leading to the change of the low-frequency behavior [Eq.~\eqref{eq:sigma_small_T>0}] and appearance of the additional peak around $\omega\sim T$ (see Fig.~\ref{fig:sigma_T>0}).
In Sec.~\ref{sec:Drude}, we investigated the spin Drude weight $D_s^\mathrm{D}$ at low temperature by computing the spectral weight arising from the regular part [Eqs.~\eqref{eq:eta}--\eqref{eq:eta_T}] and we found $D_s^\mathrm{D}>0$.
This indicates that the two-component BEC exhibits spin-superfluid characteristics with a zero spin resistivity as a superconductor has a zero electrical resistivity.
Furthermore, we showed that within the Bogoliubov theory the spin Drude weight equals the spin superfluid weight at $T\geq0$.
This suggests that at $T=0$ the spectral weight of the regular part becomes proportional to the Andreev-Bashkin drag density [Eq.~\eqref{eq:D_s^reg-rho_updown_T=0}].
Therefore, the optical spin conductivity can be regarded as a probe for the Andreev-Bashkin effect.

Regarding future works on the optical spin conductivity of Bose mixtures, there are several directions.
The first one is extension of our study to a mixture in a harmonic trap potential.
In such a situation, the peak in the spin conductivity spectrum would be shifted at a finite frequency~\cite{anderson2019conductivity}.
Comparison of the extended results with spin dipole oscillation in recent experiments~\cite{Bienaime2016spin-dipole,fava2018observation} could deepen our understanding of spin superfluidity.
Second, how the regular part and spin Drude weight change at higher temperature would be of interest.
In this case, the thermal depletion of the condensate density $n_0$ cannot be neglected and damping of quasiparticles may be important~\cite{beliaev1958energy,pitevskii1997landau,giorgini1998damping}.
In the context of connection to the Andreev-Bashkin effect, it would be important to clarify whether the equivalence between the Drude and superfluid weights of spin survives or not beyond the Bogoliubov theory.
It would be also interesting to investigate optical spin conductivity in the presence of optical lattices~\cite{jessen1996optical,krutitsky2016ultracold,Schafer2020tools} and in the quantum droplet phase~\cite{bottcher2021new,luo2020anew}.

\begin{acknowledgments}
We thank Ippei Danshita, Kazuya Fujimoto, Satoshi Fujimoto, Tomoya Hayata, Yoshimasa Hidaka, Norio Kawakami, and Takeshi Mizushima for stimulating discussions.
YS also thank Gordon Baym for enlightening comments, Donato Romito for information on longitudinal current responses within the Bogoliubov theory,
and Grigory Efimovich Volovik for communication on the Andree-Bashkin effect in liquid $^3$He.
We acknowledge JSPS KAKENHI for Grants (JP18H05406, 19J01006, JP21K03436, JP22K13981). 
YS is supported by Pioneering Program of RIKEN for Evolution of Matter in the Universe
(r-EMU).
SU is supported by MEXT Leading Initiative for Excellent Young Researchers (Grant No. JPMXS0320200002) and Matsuo Foundation.
\end{acknowledgments}

\appendix
\section{\label{appendix:depletion}Depletion of the condensate}
Here, we show that the difference of the condensate density $n_0$ from the total density $n$ is negligibly small for a weakly interacting mixture at low temperature, leading to $n_0\approx n$.
To this end, We compute
\begin{align}\label{eq:eta_dep}
\eta_\mathrm{dep}=\frac{n-n_0}{n}=\frac{1}{N}\sum_{\k\neq\0}n_\k,
\end{align}
where $n_\k=\sum_{\tau=\up,\down}\<a_{\k,\tau}^\+a_{\k,\tau}\>$ is the momentum distribution.
Using the transformation of $a_{\k,\tau}$in Eq.~\eqref{eq:b}, we obtain
\begin{align}
\eta_\mathrm{dep}
&=\sum_{\alpha=d,s}( \eta_{\mathrm{dep},\alpha}^\mathrm{Q}+ \eta_{\mathrm{dep},\alpha}^\mathrm{T}),
\end{align}
where the first term independent of $T$ provides quantum depletion
\begin{align}
\eta_{\mathrm{dep},\alpha}^\mathrm{Q}&=\frac{1}{n}\int\frac{d\k}{(2\pi)^3}\,v_{k,\alpha}^2
=\frac{8}{3\sqrt\pi}\sqrt{na_\alpha^3},
\end{align}
and the latter term denotes thermal depletion
\begin{align}
\eta_{\mathrm{dep},\alpha}^\mathrm{T}
&=\frac{1}{n}\int\frac{d\k}{(2\pi)^3}\,(u_{k,\alpha}^2+v_{k,\alpha}^2)f_\mathrm{B}(E_{\k,\alpha}).
\end{align}
At low temperature $\eta_{\mathrm{dep},\alpha}^\mathrm{T}$ is expanded as
\begin{align}
\eta_{\mathrm{dep},\alpha}^\mathrm{T}
&=\frac{2\sqrt{\pi^3}}{3}\sqrt{na_\alpha^3}\left(\frac{T}{g_\alpha n}\right)^2
\end{align}
for $T\ll g_\alpha n$ or
\begin{align}\label{eq:eta^T_g=0}
\eta_{\mathrm{dep},\alpha}^\mathrm{T}
=2\left(\frac{T}{T_\mathrm{BEC}^0}\right)^{3/2}
\end{align}
for $g_\alpha n \ll T \ll T_\mathrm{BEC}^0$.
From Eqs.~\eqref{eq:eta_dep}--\eqref{eq:eta^T_g=0}, we can find $\eta_\mathrm{dep}\ll1$ for a weakly interacting mixture ($\sqrt{na_\alpha^3}\ll1$) at sufficiently low temperature ($T\ll\mathrm{max}\{g_d n,\,g_s n\},\,T_\mathrm{BEC}^0$), leading to $n_0\approx n$.

\section{\label{appendix:superfluid_weights}Superfluid weights}
Here, we discuss the superfluid weights and relate $D_s^\mathrm{SF}$ to the Andreev-Bashkin drag density.
It is worth clarifying the difference of the superfluid weights between mass (or particle-number) and spin transport as well as the difference from the Drude weights.
In order to define these weights in a unified manner, we introduce 
\begin{align}\label{eq:D_lambda(q_T,q_L,omega)}
D_\lambda(\q_\mathrm{T},\q_\mathrm{L},\omega)&=D_0+\pi\Re\,\chi_{\lambda\lambda}(\q,\omega),
\end{align}
where $\lambda$ labels particle-number ($\lambda=n$) and spin ($\lambda=s$) degrees of freedom and $\q_\mathrm{T}=(0,q_y,q_z)$ and $\q_\mathrm{L}=(q_x,0,0)$.
The momentum-resolved current response function is 
\begin{align}\label{eq:chi_qomega}
\chi_{\lambda\lambda}(\q,\omega)&=\frac{-i}{V}\int_0^\infty dt\,e^{i\omega^+ t}\<[\tilde{j}_{\lambda,x}(\q,t),\tilde{j}_{\lambda,x}^\+(\q,0)]\>,
\end{align}
where 
$\tilde{\bm{j}}_{n/s}(\q)=\sum_\k\frac{2\k+\q}{2m}(a_{\k,\up}^\+a_{\k+\q,\up}\pm a_{\k,\down}^\+a_{\k+\q,\down})$ is a current density operator in momentum space\footnote{The spin current response function contributing to $\sigma_s(\omega)$ in Eq.~\eqref{eq:sigma} is expressed as $\chi_{ss}(\omega)=\chi_{ss}(\0,\omega)$.}.
The Drude weight $D_\lambda^\mathrm{D}$ and superfluid weight $D_\lambda^\mathrm{SF}$ are given by taking the zero-momentum and zero-frequency limits of $D_\lambda(\q_\mathrm{T},\q_\mathrm{L},\omega)$ in different order:
\begin{align}\label{eq:D_lambda^D}
D_\lambda^\mathrm{D}
&=\lim_{\omega\to0}D_\lambda(\0,\0,\omega),\\
\label{eq:D_lambda^SF}
D_\lambda^\mathrm{SF}
&=\lim_{\q_\mathrm{T}\to\0}D_\lambda(\q_\mathrm{T},\0,0),
\end{align}
while the other order of taking limits exactly provides
\begin{align}\label{eq:D_lambda^L}
\lim_{\q_\mathrm{L}\to\0}D_\lambda(\0,\q_\mathrm{L},0)=0,
\end{align}
which reflects the sum rules of density and spin structure factors~\cite{pitaevskii2003,romito2021linear}.
Equation~\eqref{eq:D_lambda^D} is consistent with Eq.~\eqref{eq:D_s0}.
For a homogeneous mixture of BECs, the Drude weight for mass is identical with the total spectral weight for any temperature and interaction strengths:
\begin{align}\label{eq:D_n^D}
D_n^\mathrm{D}=D_0.
\end{align}
This result is called Kohn's theorem and comes from the fact that $\tilde{\bm{j}}_{n}(\0,t)=\bm{P}/m$ with the total momentum $\bm{P}$ is conserved due to the translational invariance.

In the case of a binary mixture of BECs, the superfluid weights $D_n^\mathrm{SF}$ and $D_s^\mathrm{SF}$ are related to mass densities in the three-fluid hydrodynamics.
To confirm this, we follow the formalism in Ref.~\cite{romito2021linear}.
In the hydrodynamic limit, the mass current densities $m\bm{j}_\tau$ with $\tau=\up,\down$ are given by
\begin{subequations}\label{eq:three-fluid}
\begin{align}
m\bm{j}_\up&=\rho_\up^\mathrm{NF}\bm{v}^\mathrm{NF}+\rho_{\up\up}\bm{v}_{\up}^\mathrm{SF}+\rho_{\up\down}\bm{v}_{\down}^\mathrm{SF},\\
m\bm{j}_\down&=\rho_\down^\mathrm{NF}\bm{v}^\mathrm{NF}+\rho_{\down\up}\bm{v}_{\up}^\mathrm{SF}+\rho_{\down\down}\bm{v}_{\down}^\mathrm{SF},
\end{align}
\end{subequations}
where $\bm{v}^\mathrm{NF}$ and $\bm{v}_{\tau}^\mathrm{SF}$ are normal-fluid and superfluid velocities, respectively, and $\rho_\tau^\mathrm{NF}$ and $\rho_{\tau\tau'}$ are mass densities of the normal fluid and superfluids, respectively.
The off-diagonal terms $\rho_{\up\down}=\rho_{\down\up}$ of the superfluid density matrix denote the Andreev-Bashkin drag density, which determines the contribution of the superfluid velocity of one component to the mass current of the other.
In the homogeneous case, the invariance of Eqs.~\eqref{eq:three-fluid} under the Galilean transformation leads to the mass relation $m n_\tau=\rho_\tau^\mathrm{NF}+\rho_{\tau\tau}+\rho_{\up\down}$.

By using the linear response theory, the densities $\rho_\tau^\mathrm{NF}$ and $\rho_{\tau\tau'}$ are related to current response functions as follows~\cite{romito2021linear}:
\begin{subequations}\label{eq:rho_tautau'}
\begin{align}\label{eq:rho_updown0}
\rho_{\up\down}&=m^2\chi_{\up\down}^\mathrm{T},\\
\rho_{\tau\tau}&=mn_\tau+m^2\chi_{\tau\tau}^\mathrm{T},\\
\label{eq:rho_tau^NF}
\rho_{\tau}^\mathrm{NF}&=-m^2\sum_{\tau'=\,\up,\down}\chi_{\tau\tau'}^\mathrm{T},
\end{align}
\end{subequations}
where $n_\tau$ is the number density of the $\tau=\,\up,\down$ component, $\chi_{\tau\tau'}^\mathrm{T}\equiv\lim_{\q_\mathrm{T}\to0}\chi_{\tau\tau'}(\q_\mathrm{T},0)$, and
$\chi_{\tau\tau'}(\q,\omega)=\frac{-i}{V}\int_0^\infty dt\,e^{i\omega^+ t}\<[\tilde{j}_{\tau,x}(\q,t),\tilde{j}_{\tau',x}^\+(\q,0)]\>$ with $\tilde{\bm{j}}_{\tau}(\q)=\sum_\k\frac{2\k+\q}{2m}a_{\k,\tau}^\+a_{\k+\q,\tau}$.
By using $\chi_{nn/ss}=\chi_{\up\up}+\chi_{\down\down}\pm2\chi_{\up\down}$ and Eqs.~\eqref{eq:D_lambda(q_T,q_L,omega)} and \eqref{eq:rho_tautau'}, the superfluid weights in Eq.~\eqref{eq:D_lambda^SF} are found to be
\begin{subequations}\label{eq:D^SF}
\begin{align}\label{eq:D_n^SF}
D_n^\mathrm{SF}
&=\frac{\pi}{m^2}\left(mn-\rho^\mathrm{NF}\right),\\
\label{eq:D_s^SF0}
D_s^\mathrm{SF}
&=\frac{\pi}{m^2}\left(mn-\rho^\mathrm{NF}-4\rho_{\up\down}\right),
\end{align}
\end{subequations}
where $n=n_\up+n_\down$ and $\rho^\mathrm{NF}=\rho_\up^\mathrm{NF}+\rho_\down^\mathrm{NF}$.
Equations~\eqref{eq:D^SF} show that $D_n^\mathrm{SF}$ is independent of the drag density, while $D_s^\mathrm{SF}$ is affected by $\rho_{\up\down}$.
Equation~\eqref{eq:D_s^SF0} is Eq.~\eqref{eq:D_s^SF_rho}.

Next, we confirm the meaning of $D_m^\mathrm{SF}$ and $D_s^\mathrm{SF}$ in the hydrodynamic picture.
The hydrodynamic relations in Eqs.~\eqref{eq:three-fluid} can be rewritten in $\lambda=n,s$ basis as
\begin{subequations}\label{eq:three-fluid2}
\begin{align}
m\bm{j}_{n}&=\rho^\mathrm{NF}\bm{v}^\mathrm{NF}+\rho_{nn}\bm{v}_n^\mathrm{SF}+\rho_{ns}\bm{v}_s^\mathrm{SF},\\
m\bm{j}_{s}&=\rho_s^\mathrm{NF}\bm{v}^\mathrm{NF}+\rho_{sn}\bm{v}_{n}^\mathrm{SF}+\rho_{ss}\bm{v}_{s}^\mathrm{SF},
\end{align}
\end{subequations}
where 
$\bm{j}_{n/s}=\bm{j}_\up\pm\bm{j}_\down$, $\rho_s^\mathrm{NF}=\rho_\up^\mathrm{NF}-\rho_\down^\mathrm{NF}$, and $\bm{v}_{n/s}^\mathrm{SF}=(\bm{v}_\up^\mathrm{SF}\pm\bm{v}_\down^\mathrm{SF})/2$.
The superfluid weights correspond to the diagonal components of superfluid densities,
\begin{subequations}
\begin{align}
\rho_{nn}&=\frac{m^2 D_n^\mathrm{SF}}{\pi}=mn-\rho^\mathrm{NF},\\
\rho_{ss}&=\frac{m^2 D_s^\mathrm{SF}}{\pi}=mn-\rho^\mathrm{NF}-4\rho_{\up\down},
\end{align}
\end{subequations}
while the off-diagonal ones are $\rho_{ns}=\rho_{sn}=\rho_{\up\up}-\rho_{\down\down}$.
In the spin balanced case with $\rho_s^\mathrm{NF}=\rho_{ns}=0$, $\bm{j}_n$ and $\bm{j}_s$ are decoupled~\cite{kim2020obvervation}: $m\bm{j}_{n}=\rho^\mathrm{NF}\bm{v}^\mathrm{NF}+\rho_{nn}\bm{v}_n^\mathrm{SF}$ and $m\bm{j}_{s}=\rho_{ss}\bm{v}_{s}^\mathrm{SF}$.

\section{\label{appendix:current-response}Current response functions within the Bogoliubov theory}
Here, we calculate $\chi_{\lambda\lambda}(\q,\omega)$ in Eq.~\eqref{eq:chi_qomega} for the 
$\mathbb{Z}_2$ symmetric mixture 
 within the Bogoliubov theory.
As in the case of $\chi_{ss}(\omega)=\chi_{ss}(\q=\0,\omega)$ in Sec.~\ref{sec:optical_spin_conductivity}, we can straightforwardly perform the computation.
By using $a_{\k=\0,\tau}=\sqrt{N/2}$ and the Bogoliubov transformations for finite momentum [Eq.~\eqref{eq:b}], $\tilde{\bm{j}}_{n/s}(\q)=\sum_\k\frac{\k+\k'}{2m}(a_{\k,\up}^\+a_{\k',\up}\pm a_{\k,\down}^\+a_{\k',\down})$ with $\k'=\k+\q$
can be expressed in terms of quasiparticle operators $b_{\k,\alpha}$.
Substituting the obtained expressions into Eq.~\eqref{eq:chi_qomega}, we find
\begin{align}\label{eq:chi_lambda}
\chi_{\lambda\lambda}(\q,\omega)=\chi_{\lambda\lambda}^{(0)}(\q,\omega)+\chi_{\lambda\lambda}^{(-)}(\q,\omega)+\chi_{\lambda\lambda}^{(+)}(\q,\omega),
\end{align}
where
\begin{widetext}
\begin{align}\label{eq:chi_lambda^(0)}
\chi_{\lambda\lambda}^{(0)}(\q,\omega)
&=\frac{n}{2m}\frac{(q_x)^2}{\q^2}\qty(\frac{E_{q,\beta}}{\omega^+-E_{q,\beta}}-\frac{E_{q,\beta}}{\omega^++E_{q,\beta}}),\\
\label{eq:chi_lambda^(-)}
\chi_{\lambda\lambda}^{(-)}(\q,\omega)&=\frac1V\sum_{\k\neq\0,-\q}\sum_{\alpha=d,s}\qty(\frac{k_x+k'_x}{2m})^2\left[\frac{(u_{k,\alpha} u_{k',\alpha'})^2-u_{k,\alpha}u_{k',\alpha'}v_{k,\alpha}v_{k',\alpha'}}{\omega^+-(E_{k',\alpha'}-E_{k,\alpha})}-\frac{(v_{k,\alpha} v_{k',\alpha'})^2-u_{k,\alpha}u_{k',\alpha'}v_{k,\alpha}v_{k',\alpha'}}{\omega^++(E_{k',\alpha'}-E_{k,\alpha})}\right]\nonumber\\
&\qquad\qquad\qquad\qquad\times\qty[f_\mathrm{B}(E_{k,\alpha})-f_\mathrm{B}(E_{k',\alpha'})],\\
\label{eq:chi_lambda^(+)}
\chi_{\lambda\lambda}^{(+)}(\q,\omega)&=\frac1V\sum_{\k\neq\0,-\q}\sum_{\alpha=d,s}\qty(\frac{k_x+k'_x}{2m})^2\left[\frac{(v_{k,\alpha} u_{k',\alpha'})^2-u_{k,\alpha}u_{k',\alpha'}v_{k,\alpha}v_{k',\alpha'}}{\omega^+-(E_{k',\alpha'}+E_{k,\alpha})}-\frac{(u_{k,\alpha} v_{k',\alpha'})^2-u_{k,\alpha}u_{k',\alpha'}v_{k,\alpha}v_{k',\alpha'}}{\omega^++(E_{k',\alpha'}+E_{k,\alpha})}\right]\nonumber\\
&\qquad\qquad\qquad\qquad\times\qty[1+f_\mathrm{B}(E_{k,\alpha})+f_\mathrm{B}(E_{k',\alpha'})].
\end{align}
\end{widetext}
Here, the label $\beta$ takes $\beta=d$ for $\lambda=n$ ($\beta=s$ for $\lambda=s$).
In the case of $\chi_{nn}(\q,\omega)$, $\alpha'$ is identical with $\alpha$ ($\alpha'=\alpha$).
On the other hand, for the spin current response $\chi_{ss}(\q,\omega)$, $\alpha'$ denotes the mode opposite to $\alpha$.
Taking $\q=\0$ in Eqs.~\eqref{eq:chi_lambda}--\eqref{eq:chi_lambda^(+)}, we can reproduce Eq.~\eqref{eq:chi1}.

Equations~\eqref{eq:chi_lambda}--\eqref{eq:chi_lambda^(+)} for $\chi_{ss}(\q,\omega)$ show that the transverse limit and zero-frequency limit are commutable:
\begin{align}\label{eq:chi_ss^T}
    &\Re\,\chi_{ss}(\q_\mathrm{T}\to\0,0)=\Re\,\chi_{ss}(\0,\omega\to+0)\nonumber\\
    &=-\frac{4}{3mV}\sum_{\k\neq\0}\frac{(g_{\up\down}n)^2(\epsilon_k)^3}{E_d(\epsilon_k)E_s(\epsilon_k)} 
    \sum_{\nu=\pm}
    \frac{F_\nu(\epsilon_k)}{[E_\nu(\epsilon_k)]^3},
\end{align}
where we used $E_{k,d}\neq E_{k,s}$ in the presence of the intercomponent interaction ($g_{\up\down}\neq0$).
By combining this with Eqs.~\eqref{eq:D_lambda(q_T,q_L,omega)}, \eqref{eq:D_lambda^D}, and \eqref{eq:D_lambda^SF}, our results with the Bogoliubov theory suggest that the Drude and superfluid weights for spin are identical:
\begin{align}
    D_s^\mathrm{D}=D_s^\mathrm{SF}.
\end{align}

In contrast to the spin current response, $\chi_{nn}(\q,\omega)$ is sensitive to the order of taking $\q_\mathrm{T}\to\0$ and $\omega\to+0$:
\begin{align}\label{eq:chi_nn^T}
    \Re\,\chi_{nn}(\q_\mathrm{T}\to\0,0)&=\frac1V\sum_{\k\neq\0}\sum_{\alpha}\frac{k^2}{3m^2}\pdv{f_\mathrm{B}(E_{k,\alpha})}{E_{k,\alpha}},\\
    \label{eq:chi_nn^omega}
    \Re\,\chi_{nn}(\0,\omega\to+0)&=0.
\end{align}
This results from the fact that $E_{k',\alpha'}$ in Eqs.~\eqref{eq:chi_lambda^(-)} and \eqref{eq:chi_lambda^(+)} approaches $E_{k,\alpha}$ in the limit of $\q\to\0$ for the mass current response with $\alpha'=\alpha$.
This sensitivity to the order of taking limits is physically reasonable because Eq.~\eqref{eq:chi_nn^T} together with Eq.~\eqref{eq:rho_tau^NF} leads to the normal fluid density~\cite{fil2005nondissipative}
\begin{align}\label{eq:rho^NF_Bogoliubov}
    \rho^\mathrm{NF}
    &=-\frac{2m}{3V}\sum_{\k\neq\0}\sum_{\alpha}\epsilon_k\pdv{f_\mathrm{B}(E_{k,\alpha})}{E_{k,\alpha}},
\end{align}
which is finite at $T>0$, while Eq.~\eqref{eq:chi_nn^omega} is consistent with the statement of the Kohn's theorem [Eq.~\eqref{eq:D_n^D}].

We next rederive the drag density $\rho_{\up\down}$ for a $\mathbb{Z}_2$ symmetric mixture.
In terms of spin and mass current responses, the drag density in Eq.~\eqref{eq:rho_updown0} is given by $\rho_{\up\down}=\frac{m^2}{4}[\Re\,\chi_{nn}(\q_\mathrm{T}\to\0,0)-\Re\,\chi_{ss}(\q_\mathrm{T}\to\0,0)]$.
Substituting Eqs.~\eqref{eq:chi_ss^T} and \eqref{eq:chi_nn^T} into this yields~\cite{fil2005nondissipative}
\begin{align}
    \rho_{\up\down}
    &=\frac{m}{3V}\sum_{\k\neq\0}\frac{(g_{\up\down}n)^2(\epsilon_k)^3}{E_d(\epsilon_k)E_s(\epsilon_k)} 
    \sum_{\nu=\pm}
    \frac{F_\nu(\epsilon_k)}{[E_\nu(\epsilon_k)]^3}\nonumber\\
    &\quad+\frac{m}{6V}\sum_{\k\neq\0}\sum_{\alpha}\epsilon_k\pdv{f_\mathrm{B}(E_{k,\alpha})}{E_{k,\alpha}}.    
\end{align}
In particular, the drag density at $T=0$ is
\begin{subequations}\label{eq:rho_updown}
\begin{align}
\rho_{\up\down}&=\frac{mn}{2\sqrt{2}}\sqrt{na^3}z^2\mathcal{F}(z),\\
\mathcal{F}(z)&=\frac{256}{45\sqrt{2\pi}}\frac{2+3\sqrt{1-z^2}}{[\sqrt{2(1-z)}+\sqrt{2(1+z)}]^3}
\end{align}
\end{subequations}
with $z=a_{\up\down}/a$.

We finally note the longitudinal current responses within the Bogoliubov theory.
As pointed out in Ref.~\cite{romito2021linear}, it has the shortcoming such that
the exact sum rules are explicitly broken.
Such a shortcoming should be cured by considering vertex corrections in a similar manner to superconductors~\cite{schrieffer1964theory}.

\bibliography{masterbib.bib}
\end{document}